\documentclass[12pt]{iopart}

\usepackage{graphicx}
\usepackage{amsfonts}
\usepackage{psfrag}

\def\be{\begin{equation}}
\def\ee{\end{equation}}
\def\bea{\begin{eqnarray}}
\def\eea{\end{eqnarray}}

\newcommand{\kB}{k_{\scriptscriptstyle \rm B}}
\newcommand{\Tc}{T_{\scriptstyle \rm d}}

\newcommand{\qEA}{q_{\scriptscriptstyle \rm EA}}
\newcommand{\pEA}{p_{\scriptscriptstyle \rm EA}}
\newcommand{\chith}{\chi_{\scriptstyle \rm th}}
\newcommand{\chihet}{\chi_{\scriptstyle \rm het}}
\newcommand{\Fth}{{\cal F}_{\scriptstyle \rm th}}
\newcommand{\Fhet}{{\cal F}_{\scriptstyle \rm het}}

\begin{document}

\title{Finite-size critical fluctuations in microscopic models of
  mode-coupling theory}

\author{Silvio Franz}

\address{Laboratoire de Physique Th\'eorique et Mod\`eles
  Statistiques, \\ CNRS et Universit\'e Paris-Sud 11, UMR8626,
  B\^at. 100, 91405 Orsay Cedex, France}
\ead{silvio.franz@lptms.u-psud.fr}

\author{Mauro Sellitto}

\address{Dipartimento di Ingegneria Industriale e dell'Informazione,
  \\ Seconda Universit\`a di Napoli, Real Casa dell'Annunziata, I-81031
  Aversa (CE), Italy} \ead{mauro.sellitto@unina2.it}

\begin{abstract} 
 Facilitated spin models on random graphs provide an ideal microscopic
 realization of the mode-coupling theory of supercooled liquids: they
 undergo a purely dynamic glass transition with no thermodynamic
 singularity. In this paper we study the fluctuations of dynamical
 heterogeneity and their finite-size scaling properties in the $\beta$
 relaxation regime of such microscopic spin models. We compare the
 critical fluctuations behavior for two distinct measures of
 correlations with the results of a recently proposed field
 theoretical description based on quasi-equilibrium ideas. We find
 that the theoretical predictions perfectly fit the numerical
 simulation data once the relevant order parameter is identified with
 the persistence function of the spins. 
\end{abstract}

\maketitle

\section{Introduction}

Facilitated spin models (FSM) have been proposed as an elegant way to
rationalize the idea that the glass transition is purely kinetic in
nature~\cite{FrAn}. This contrasts the view that glassy phenomena in
liquids can be understood in terms of a rugged free-energy landscape,
as it happens in systems with complicated static interactions. Recent
results have shown, however, that a common mechanism could be at the
basis of the slowing down of relaxation both in FSM and in landscape
dominated systems. Important insight has come from the study of
cooperative FSM on random graphs, as they provide a suitable mean
field limit in which the possibly sharp dynamic crossover observed in
finite-size systems becomes an actual transition in the thermodynamic
limit. In \cite{mauroetal,ArSe2012} it has been shown that
relaxational dynamics of FSM on random graphs verifies the universal
relations predicted by Mode Coupling Theory (MCT) of supercooled
liquids~\cite{Goetze,Patrick}. In fact, the formal structure of MCT
singularities is closely related to that appearing in the bootstrap
percolation analysis of FSM~\cite{Se2012} and $k$-core problems on
random graphs. In some cases, it is known that rugged landscapes and
kinetic constraints provide a dual description of slow relaxation
\cite{newman, garrahan}.  The recently studied case of the random
{\small XOR-SAT} optimization problem of information theory~\cite{FKZ} 
explicitly shows the emergence of a random first order transition
scenario and the highly nonlocal nature of the relation between such
dual descriptions.  Unfortunately, in the general case, a detailed
analytic comprehension of the long time dynamics is lacking. Despite
the bootstrap percolation analysis allows to compute the phase diagram
and the arrested part of correlation, important characteristics of
relaxation, like the origin of the feedback mechanism leading to MCT
behavior and the prediction of relaxation exponents remain out of
reach. On general ground, MCT describes dynamical arrest as a critical
phenomenon driven by long lived dynamic heterogeneities with
increasing correlations.  Within the mean-field scenario this
transition has universal characteristics that can be rationalized
postulating random critical temperature variations due to disorder
\cite{Sarlat}. In the so-called $\beta$ relaxation regime -- i.e. the
time window in which the correlation function approaches its plateau
value corresponding to the glass transition and the system explores
the configuration space near the ideal arrested state -- it is
possible to derive from first principle the postulated random critical
point, and to describe overlap fluctuations in terms of a cubic field
theory in a random field \cite{gangof4}. This theory, initially
formulated for systems with non trivial Hamiltonian like spin glasses
and liquids, identifies the overlap with the initial condition as the
order parameter in terms of which a field theoretical description of
fluctuations is possible.

In this context it is natural to ask whether FSM conform to the
expectations of this quasi-equilibrium field theory. The problem is
not obvious as metastability and dynamic arrest in FSM are induced by
purely kinetic constraints and the Hamiltonian is generally trivial
(there is neither quenched disorder nor frustrated {\em static}
interaction in the problem). As we shall see, the overlap fails to
specify long lived metastable states and one has therefore to turn to
other possible correlation functions capable to characterize
metastable states.  The correspondence with the bootstrap percolation
problem in fact suggests to use the core and its time dependent
analogue, the persistence function, as an appropriate order parameter
for the glass transition. In this paper we study numerically the
finite size properties of both the fluctuations of the persistence
function and the spin overlap. We find that while the former follows
scaling laws with the exponents suggested by the cubic random field
theory, the scaling of the overlap fluctuations does not.

The scheme of the paper is the following. In the second section we
briefly review the theory of \cite{gangof4}.  In the third section we
introduce the model and compare the numerical simulation results to
the theory predictions for two different measures of correlations.  In
the last section we discuss the main implications of our findings and
conclude the paper.

\section{Field-theoretical approach to fluctuations in the $\beta$
  regime: a mini-review}

The theory proposed in \cite{gangof4} deals with equilibrium
fluctuations of the local overlap, $q_x(s,s_0)$, between an initial
configuration, $s_0=s(0)$, and the one visited by the system at a
certain time $t$ during its evolution, $s=s(t)$.  The appropriate
definition of overlap may be system dependent and some coarse graining
is generally assumed. For example, in liquid systems one can divide
the space in small cells centered around the position $x$ and compare
the local particle density between the current and the initial
configuration. In spin systems the most natural definition is the
usual scalar product between the two spin configurations.

The theory suggests to eliminate the temporal dependence of the
relevant physical quantities from the description, starting from the
observation that due to time scale separation, when the dynamical MCT
transition is approached, dynamics can be described as a random walk
between metastable states, equilibration within a state establishes
before a new state is found. In such conditions one can eliminate time
in favour of the local overlap with the initial state and describe
fluctuations in a restricted equilibrium ensemble with probability
measure:
\begin{eqnarray}
  \mu(s|s_0)=\frac{1}{Z[s_0]}\exp\left( -\beta {\cal H}(s)+\int \!\!
  dx \, \nu(x) \, q_x(s,s_0) \right) ,
\label{meas}
\end{eqnarray} 
where ${\cal H}(s)$ is the system Hamiltonian and $\beta=1/\kB T$ is
the inverse temperature. The Lagrange multiplier $\nu(x)$ is fixed in
such a way that the average values of the local overlap $q_x(s,s_0)$
take some prescribed values close to the plateau value, $\qEA$.  The
crucial assumption in the definition of the restricted Boltzmann-Gibbs
measure, Eq.~(\ref{meas}), is that the overlap provides a sufficient
determination of the metastable states.

A deep analysis of such a theory close to the ideal MCT transition
(when activated processes are neglected) shows the emergence of a
universal description in terms of the local overlap fluctuation
\begin{eqnarray}
  \phi(x) = q_x(s,s_0)-\qEA,
\end{eqnarray} 
which is encoded in the effective action:
\begin{eqnarray}
  {\cal S}[\phi]=\int \! dx \, \left[ \frac{1}{2} \left( \nabla \phi
    \right)^2 + \epsilon \phi(x) + g \phi(x)^3 + h(x) \phi(x) \right],
\label{field}
\end{eqnarray} 
where $\epsilon$ is a parameter quantifying the distance from the
glass state (e.g., $\epsilon = T-\Tc$) and $g$ is a coupling
constant. The ``external field'' $h(x)$ account for the effects of
heterogeneity in the initial condition: it is a random Gaussian
variable with zero mean and variance
\begin{eqnarray}
{E}(h(x)h(y)) &=& \Delta \,\delta(x-y),
\end{eqnarray} 
where $\Delta$ is a system dependent parameter.  The effective action
can be used to study the finite-size corrections in mean-field glassy
systems. In that case one only deals with global overlap fluctuations
and so one just needs to replace the space integral with an overall
volume factor $N$. The field variance $\Delta$ scales then as $1/N$
and the Gaussian theory suffices to understand the finite-size scaling
properties.
 
In order to characterize fluctuations two types of four-point
correlation functions can be introduced:
\begin{eqnarray}
  \chith(t) &=& N \left[ \overline{\langle \phi(t)^2 \rangle} -
    \overline{\langle \phi(t) \rangle^2} \right] \,, \\ 
\chihet(t) &=& N
  \left[ \overline{ \langle \phi(t)\rangle^2} - \overline{\langle
      \phi(t) \rangle}^2 \right] \,.
\end{eqnarray}
We denote by the angular brackets, $\langle \cdots \rangle$, the
average over trajectories that start from the same initial
condition. This was called iso-configurational average
in~\cite{propensity1, propensity2}. In our stochastic dynamics the
iso-configurational average represents the average over the thermal
noise along the trajectories. The initial condition, denoted by $s(0)
= s_0$, will always be chosen as an equilibrium configuration and the
corresponding average will be denoted by the overbar
$\overline{\cdots}$. Thus, $\chith(t)$ characterizes thermal
fluctuations for fixed initial condition, while $\chihet(t)$
characterizes the fluctuations with respect to the initial condition.
The sum of these two functions gives back the total order parameter
fluctuations
\begin{equation}
  \chi_4(t) = N \left[ \overline{\langle \phi(t)^2 \rangle} -
    \overline{ \langle \phi(t) \rangle}^2 \right] .
\end{equation}
As typical in random field systems $\chith$ and $\chihet$ have
different scaling properties. If the quasi-equilibrium assumption of
the theory is verified one can then eliminate the explicit time
depedence of fluctuations in favor of $\phi$, and it turns out that
\begin{eqnarray}
\label{chi_th}
  \chith(\phi,\epsilon,N) &=& N^{\frac{1}{4}} \ \Fth (\phi
  N^{\frac{1}{4}}, \epsilon N^{\frac{1}{2}}) \,, \\
\label{chi_het}
  \chihet(\phi,\epsilon,N) &=& N^{\frac{1}{2}} \ \Fhet (\phi
  N^{\frac{1}{4}}, \epsilon N^{\frac{1}{2}}) \,.
\end{eqnarray}
The questions that we address in this paper are the following: 
\begin{itemize}
\item[-] Can dynamical fluctuations of FSM on random graphs be
  described in terms of the effective field theory (\ref{field})?
\item[-] What is the relevant measure of correlations that
  characterizes metastability in FSM on random graphs?
\end{itemize}
Some of the basic ingredients of the theory, namely time scale
separation and quasiergodic exploration of metastable states are
manifestly true in FSM.  However, the conventional spin overlap does
not provide a sufficient determination of metastable states.  The
glass singularity of FSM is associated to a core or bootstrap
percolation transition of frozen spins. The order parameter is
therefore the fraction of frozen spins in the core. This suggests that
the persistence function, $p(t)$, i.e. the fraction of spins that have
not flipped over the time interval $[0,t]$, is the natural correlation
function that can be used to identify metastable states. Conversely,
the conventional spin overlap, $q(t)$, fails to identify metastable
states: This is best understood in the frozen phase, where if one
takes two equilibrium configurations with a value of the overlap
corresponding to $\qEA=\lim_{t\to\infty} q(t)$ and applies the
bootstrap algorithm one typically finds distinct core configurations.

Ideally, one could imagine that it is possible to derive a MCT
equation for $p(t)$, similarly to what has been done for the minimal
size rearrangements in related models for the glass
transition~\cite{MoSe}.  Even more optimistically one could hope to
write a field-theory for persistence fluctuations like
Eq.~(\ref{field}). Unfortunately, this is a nontrivial task because
$p(t)$ is a highly non time-local quantity and its theoretical
analysis is difficult. Rather than following this line of tought, in
this paper we limit ourself to test in numerical simulations the
predictions, Eqs.~(\ref{chi_th}) and (\ref{chi_het}), if one
concentrates on the persistence function and the conventional overlap.

\section{Numerical simulations of facilitated spin 
models on random graphs}

We have studied numerically the dynamics of a facilitated system of
$N$ Ising spins, $s_i= \pm1$ with $i=1,\dots, \, N$, on a Bethe
lattice with coordination number $z$.  The system has a trivial
Hamiltonian,
\begin{equation}
  {\cal H} = - \frac{h}{2} \sum_{i=1}^N s_i \,,
\end{equation}
while spins evolve according to a Metropolis-like constrained
dynamics: at each time step a randomly chosen spin, $s_i$, is flipped
with transition probability:
\begin{eqnarray}
  w(s_i \to -s_i) &=& {\rm min} \left\{ 1, {\rm e}^{- h s_i/\kB T}
  \right\},
\end{eqnarray}
if and only if at least $f$ of its $z$ neighboring spins are in the
state $-1$. Without loss of generality we set $h=\kB$ hereafter.  The
Bethe lattice geometry is known to give a mean field description of
the system, with the advantage of preserving the notion of distance
and that of finite connectivity. More precisely, Bethe lattices are
intended here as random regular graphs, i.e. graphs chosen uniformly
among the set of graphs with $N$ vertices where all sites have exactly
$z$ neighbors.  Since all sites have the same number of neighbours
there is virtually no boundary in the system and the graphs converge
locally to trees for large $N$. It is this local tree-like structure,
in fact, that allows to derive exactly some properties of FSM.

We consider here the cooperative case $f=2$ with $z=4$, which is
characterised by a discontinuous glass transition at a critical value
$\Tc \simeq 0.481$ where the long-time limit of the persistence
function, $p(t)$, jumps from zero to a finite value $\pEA \simeq
0.673$~\cite{SeBiTo}.  In full analogy with the mixed
discontinuous-continuous nature of the non-ergodicity parameter in
MCT, the persistence has a square root singularity coming from low
temperatures and share many other salient features of
MCT~\cite{SeBiTo,mauroetal,ArSe2012,Se2012}.  Microscopically, the
discontinuity corresponds to the sudden emergence at $\Tc$ of a giant
cluster of permanently frozen spins (the core) with compact structure.
From an algorithmic poit of view this phenomenon is equivalent to the
impossibility, below $\Tc$, of generating an equilibrium configuration
in which every spin is not permanently frozen.  Metastable states can
be identified with the set of spin configurations having the same core
and their geometric organization is pretty much similar to that of
{\small TAP} states in disordered $p$-spin systems~\cite{SeBiTo}.

Apart from their conceptual underpinning such models offer the
nontrivial advantage of requiring no thermal equilibration therefore
allowing the investigation of relatively large system sizes and
samples with a moderate computational effort.  Efficient simulations
are performed by means of a faster-than-the-clock
algorithm~\cite{Krauth}. We have simulated spin systems of size
ranging from $N=128$ to $N=8192$ for a number of samples of the order
about $10^3$ to $10^4$, depending on the system size.

\begin{figure} 
\input{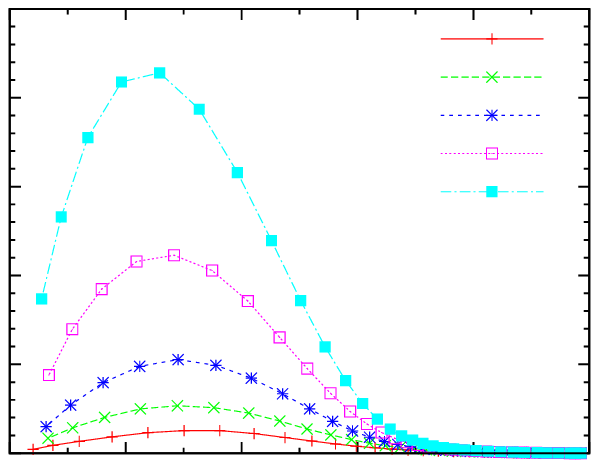}\input{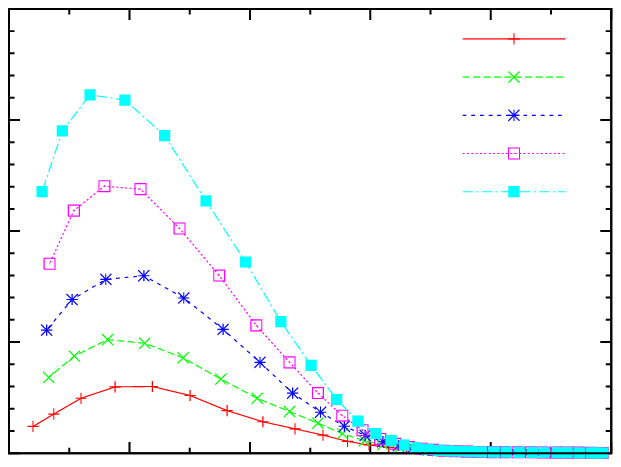}
\caption{Parametric plot of the two components of persistence
  fluctuations $\chihet(t)$ and $\chith(t)$ vs. the average persistence,
  $p(t)$, at temperature $T=0.5$ close to the glass transition ($\Tc
  \simeq 0.481$) and for various values of the system size $N$.}
\label{fig.chi4_T05}
\end{figure}

Glassy features in FSM are intimately accompanied by a substantial
growth of dynamical heterogeneity as the transition point is
approached~\cite{SeBiTo}.  Previous studies~\cite{ChChCuReSe} have
stressed the importance of the emerging time-reparametrization
invariance of soft modes fluctuations during the aging
dynamics~\cite{ChCu,Horacio,Fede}.  Other manifestations of dynamic
heterogeneity that have been studied include the Kovacs memory
effect~\cite{Kovacs} and the four-point dynamical
susceptibility~\cite{Mulet}.  We shall focus here on the system size
dependence of correlation fluctuations at criticality.  To measure
$\chihet$ and $\chith$ we generate, for each sample $\ell =1, \dots,
M$, two independent dynamical trajectories $k=1,\,2$ starting from the
same equilibrium configuration and evolving with different thermal
noise. Then for each sample and each trajectory we measure the
persistence function $\phi_{\ell,k}$, where the initial state is equal
for the two dynamical trajectories. The two components of persistence
fluctuations are thus estimated as
\begin{eqnarray}
  \chith &=& N \left[ \frac{1}{2 M} \sum_{\ell,k} \phi_{\ell,k}^2 -
    \frac{1}{M} \sum_{\ell} \phi_{\ell,1}\phi_{\ell,2} \right] \,,
  \\ \chihet &=& N \left[ \frac{1}{M} \sum_{\ell}
    \phi_{\ell,1}\phi_{\ell,2} - \left( \frac{1}{2 M} \sum_{\ell, k}
    \phi_{\ell,k} \right)^2 \right] \,.
\end{eqnarray}
Notice that since quenched disorder is absent in such FSM there is no
additional source of fluctuations apart from that one due to the
topology fluctuations in the random generation of graphs. We have
checked that such an effect is substantially negligible as expected
from the self-averaging property of random graphs.

\begin{figure} 
\input{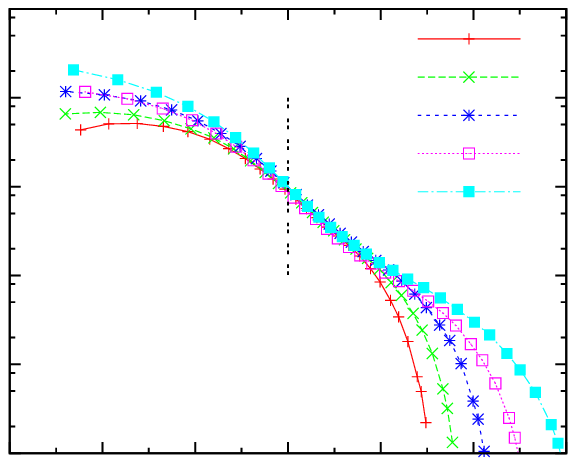}\input{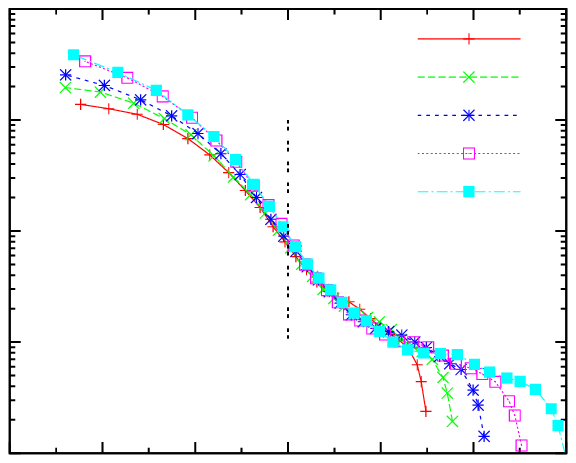}
\caption{Scaling of persistence fluctuations in the $\beta$ relaxation
  regime at the critical glass temperature $\Tc \simeq 0.481$ in the
  parametric plot representation.  The initial condition of dynamics
  is sampled from the equilibrium measure and configurations with
  permanently frozen spins are disallowed.}
\label{fig.chi4_scaling}
\end{figure}

In fig. 1 we illustrate how the two components of persistence
fluctuations at temperature slightly above $\Tc$ look like in the
quasi-equilibrium description when they are expressed in terms of the
persistence ``clock''. We observe that heterogeneity fluctuations are
almost one order of magnitude larger than thermal ones though they
present qualitatively the same behaviour. Results are similar to those
obtained in the $p$-spin spherical model.
To better mirror realistic situations in which the system size is
finite but very large we have disallowed, in the simulations of
fig. 1, initial configurations having a fraction of permanently frozen
spins.  This is reasonable as long as the dynamics is time reversible
and occurs above $\Tc$. This corresponds to the situation in which the
amorphous solid has been formed by a slow enough cooling procedure
such that irreversibility effects are negligible. In fact, permanently
frozen spins are not the results of the microscopic dynamics which
obeys the detailed balance condition. Rather, they can only be due to
``external'' influences which are ultimately determined by the
experimental conditions.  For finite systems at temperature
sufficiently close to criticality, or even below $\Tc$, the absence of
a core in a sample is an assumption that cannot be obviously
guaranteed.  For this reason we have considered two distinct
situations. In the first one, the sampling of the initial condition is
restricted to the subset of those equilibrium configurations which do
not have permanently frozen spins.  In the second situation, the
initial equilibrium configurations are sampled without such a
restriction.  In this case, regions of permanently frozen spin may
exist in a sample.  Even though they do not directly contribute to
fluctuations such frozen spins may have an indirect influence on those
spins located near the core boundary. A priori such influence might
propagate to the entire system if the core has a special structure and
is large enough. So it is important to assess the core contribution to
fluctuations scaling near criticality.

\begin{figure} 
\input{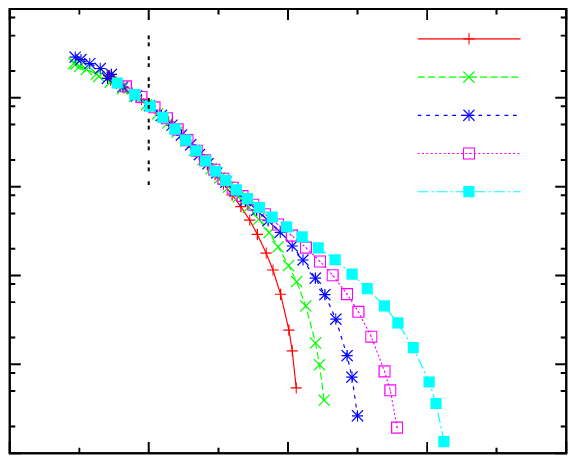}\input{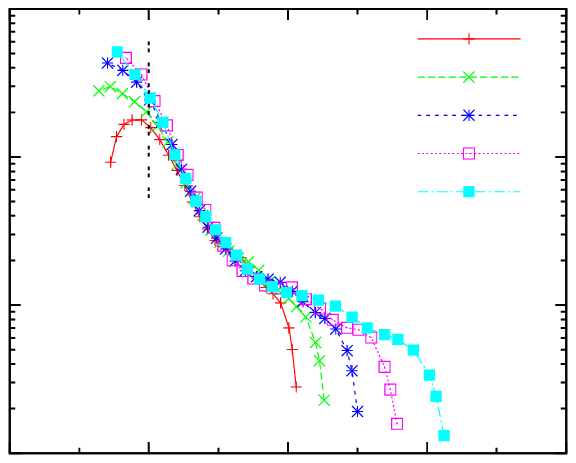}
\caption{Scaling of heterogeneity and thermal fluctuations in the
  situation in which the initial condition is sampled from the
  equilibrium measure with no restriction.}
\label{fig.neq_scaling}
\end{figure}

We have therefore studied the finite-size scaling properties of
$\chith$ and $\chihet$ at the glass transition temperature for these
two distinct situations. At $T= \Tc$ ($\epsilon=0$) the
Eqs.~(\ref{chi_th}) and (\ref{chi_het}) read
\begin{eqnarray}
\label{chi_th_0}
  \chith(\phi,0,N) &=& N^{\frac{1}{4}} \ \Fth (\phi N^{\frac{1}{4}},0) \,, \\ 
\label{chi_het_0}
  \chihet(\phi,0,N) &=& N^{\frac{1}{2}} \ \Fhet (\phi N^{\frac{1}{4}},0) \,,
\end{eqnarray}
where $\phi= p -\pEA$.  We first consider the case in which the are no
permanently frozen spin in the initial condition. We have checked that
the curves of $\chith \ N^{ - \frac{1}{4}}$ and $\chihet \ N^{ -
  \frac{1}{2}}$ as a function of $p$ obtained for various values of
$N$ cross precisely at the exactly known value of $\pEA$.  In fig.~2
we see that, in agreement with the analytical predictions, the
different curves of rescaled fluctuations plotted against the scaling
variable $N^{−\frac{1}{4}} \phi$ produces a pretty nice data collapse
in a relatively wide region around the origin.  We then consider the
situation in which the initial condition is sampled from the
equilibrium measure with no restriction.  Even though the fluctuation
curves are slightly different than the previous case we see in fig.~3
that the predicted finite-size properties are met, implying that
the core has a negligible influence on fluctuations scaling near
criticality.  We can thus summarise our first test by stating that the
theory predictions are excellently verified when the correlations are
measured in terms of the persistence function, no matter the choice of
the initial condition.

Next, we study equilibrium fluctuations of the conventional overlap,
$q(t)$, that for our system should be more properly defined as
\begin{eqnarray}
  q(t) &=& \frac{1}{1 - m^2(0)}\left[ \frac{1}{N} \sum_{i=1}^N s_i(0)
    s_i(t) - m(0) m(t) \right],
\end{eqnarray}
where $m$ is the magnetization density. Notice that $q(t)$ exhibits
features similar to the persistence function, i.e., for large system
sizes its long-time limit jumps at $\Tc$ from zero to $\qEA\simeq 0.4$
and has a square-root singularity below $\Tc$. It can play,
consequently, the legitimate role of order parameter. However, as we
have seen, this does not guarantee that metastable states are well
characterised.  Indeed, we find that two components of the spin
overlap fluctuations do not obey the predicted scaling,
Eqs.~(\ref{chi_th_0}) and (\ref{chi_het_0}). On one hand, we observe
that the size dependence of heterogeneity fluctuations, $\chihet$, is
substantially weaker than the expected $N^{\frac{1}{2}}$: a scaling
form is recovered only if $\chihet$ is rescaled by a factor
$N^{-0.35}$, rather than $N^{-\frac{1}{2}}$, see fig.~4 (left).  On
the other hand, we find that thermal fluctuations, $\chith$, are of
order one in the whole range $q > \qEA$ and do not depend on the
system size as it is observed in fig.~4 (right). These results confirm
that although the overlap can be used as an order parameter to
describe the glass transition, it yields only a partial
characterization of metastability in FSM~\cite{FKZ}. Therefore, when
the nature of metastable states in the system under consideration is
not exactly known it is a rather delicate task to give a universal
description of finite-size critical fluctuations.

\begin{figure} 
\input{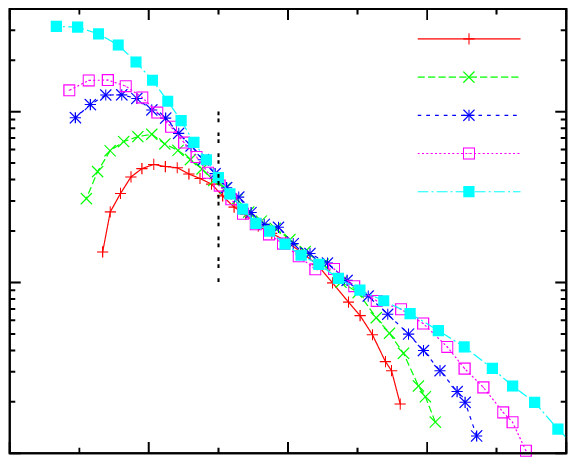}\input{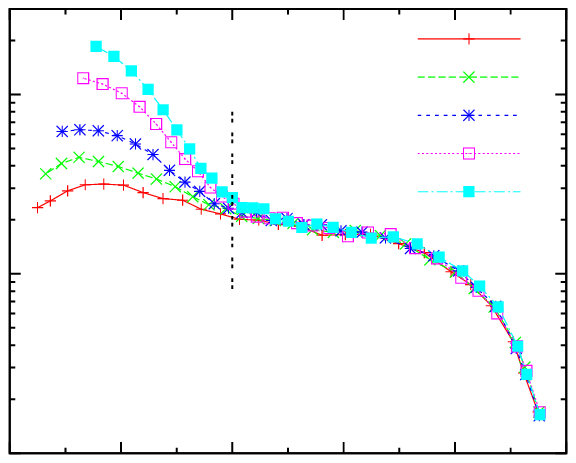}
\caption{Heterogeneity and thermal overlap fluctuations at criticality
  $T=\Tc$ in the parametric plot representation ($\qEA \simeq
  0.4$). Initial equilibrium configurations with permanently frozen
  spins are disallowed here.}
\label{fig.chi4}
\end{figure}

Finally, we note that finite-size analysis of core fluctuations near
the bootstrap percolation transition have been performed in
Refs.~\cite{DeMo,IwSa}.  These works are concerned with an
edge-deletion process (or {\it leaf removal} algorithm) that
corresponds, unlike our case, to an irreversible dynamics.  This
stochastic process is endowed with an absorbing state and is described
by a one dimensional Langevin equation in a cubic potential and
Gaussian noise. The absorbing state corresponds to the presence of a
core and the dynamics occurs, in our context, below the critical
temperature $\Tc$, i.e. in the frozen phase.  Our results for
fluctuation exponents and scaling variables are in agreement with
those found in Refs.~\cite{DeMo,IwSa} when the time dependence of the
various quantities is parametrically expressed in terms of the
persistence function.  This happens in spite of the deep difference
between two dynamical processes: irreversible for the leaf-removal
algorithm and reversible for the facilitated dynamics.  This suggests
that once the persistence function is fixed, both dynamical processes
sample the configuration space with the same law.

\section{Discussion and conclusion}

We have investigated the finite-size scaling properties of critical
fluctuations in the $\beta$ relaxation regime of a cooperative
facilitated spin model on regular random graphs for two different
correlation measures.  We tested our numerical simulation results
against the predictions of a recent field theoretical approach that
describes fluctuations in the $\beta$ regime in terms of a cubic field
theory in a random field. We find it remarkable that although FSM have
been purposely devised to lack any thermodynamical content, the
predictions of the quasi-equilibrium approach are exactly matched when
correlations are measured in terms of the persistence function.  In
contrast, when correlations are measured through the conventional
overlap function we find that theory predictions are not verified.
Such a discrepancy can be traced back to the fact that the
conventional overlap function does not provide a sufficient
characterization of metastable states in FSM. Thus, our findings
confirm the universal character of fluctuations close to ideal MCT
transitions, provided that the order parameter is correctly
identified.  In the presence of an incomplete knowledge about the very
nature of the metastability, some cautions is required when
simulations or experiments are compared with theory.

Outside the mean-field scenario, when one turns to finite dimensional
systems, the MCT transition is generally expected to become a
cross-over in which critical effects get mixed up with thermally
activated hopping processes.  In this case the notion of metastability
is not sharply defined and does require the introduction of a suitable
time-scale, possibly dependent on the observation time. Evidently, in
such a situation the appropriate choice of relevant correlations
becomes even more crucial.  For these reasons, it would certainly be
interesting to extend the present investigation to finite dimensional
FSM.  This would give the possibility to assess, on one hand, the role
of finite-size corrections to scaling, which in bootstrap percolation
related problems are known to be anomalously large and, on the other
hand, the relevance of the above scenario in systems in which the
glass transition is avoided in the infinite size limit.  Also, for
continuous glass transitions, as those appearing in the simplest
higher-order singularity scenario of MCT~\cite{Goetze,ArSe2012}, we
expect the above field theoretical description be still valid provided
that the action ${\cal S}$ includes a quartic term, and the scaling
behaviour of fluctuations is modified accordingly
\cite{FrPa-rfim,ArFrSe}.  Finally, from a more speculative theoretical
perspective, we observe that since the aging dynamics of glassy
systems can be interpreted in terms of quasi-equilibrium concepts one
would be tempted to generalise the field theoretical scenario we
tested above in a suitable defined out of equilibrium regime of lenght
and time scales.

\end{document}